\documentclass{PoS}
\pdfoutput=1

\providecommand{\rd}{\ensuremath{\mathrm{d}}}
\providecommand{\PgLb}{\ensuremath{\Lambda_\mathrm{b}}} 
\newcommand{\Lbpsilam}{\ensuremath{\PgLb\to\JPsi\Lambda}}
\newcommand{\BLbpsilam}{\ensuremath{\calB(\Lbpsilam)}}
\newcommand{\calB}{\ensuremath{\mathcal{B}}}
\newcommand{\JPsi}{\ensuremath{J\!/\!\psi}}
\newcommand{\dsdptbf}{\ensuremath{\rd\sigma/\rd p_\mathrm{T}^{\PgLb}\times\BLbpsilam}}
\newcommand{\dsdybf}{\ensuremath{\rd\sigma/\rd y^{\PgLb}\times\BLbpsilam}}
\newcommand{\psimumu}{\ensuremath{\JPsi\to\mu^+\mu^-}}

\newcommand{\lamppi}{\ensuremath{\Lambda\rightarrow p \pi^-}}

\title{Heavy-Flavor Results from CMS}

\ShortTitle{CMS Heavy Flavor}

\author{\speaker{Keith A. Ulmer}\\
        University of Colorado\\
        on behalf of the CMS Collaboration\\
        E-mail: \email{keith.ulmer@colorado.edu}}

\abstract{Heavy-flavor physics offers the opportunity to make indirect tests of physics beyond the Standard Model
through precision measurements, and of quantum chromodynamics (QCD) through particle production studies. 
The rare decays $B^0_s(B^0)\rightarrow \mu^+\mu^-$ and $D^0\rightarrow \mu^+\mu^-$ are excellent tests of the flavor
sector of the Standard Model and are sensitive to new physics. We report on studies of these decays and
present the first observation of the excited $b$ baryon $\Xi_b^{*0}$ in strong decays to $\Xi_b^{-}$ and a charged pion,
the observation of two $B_c$ meson decay modes and production properties of the $\Lambda_b$ baryon, all
performed with the CMS experiment in $pp$ collisions at $\sqrt{s}=7$ TeV.}

\FullConference{36th International Conference on High Energy Physics,\\
		July 4-11, 2012\\
		Melbourne, Australia}

\begin{document}

\section{Introduction}

The CMS experiment at the LHC pursues a diverse program of heavy-flavor measurements. We present here some recent
highlights of this program, including searches for the rare decays
$B_s(B^0)\rightarrow \mu^+\mu^-$ and $D^0\rightarrow \mu^+\mu^-$, which are excellent tests of the flavor
sector of the Standard Model with sensitivity to possible physics beyond the Standard Model, and the first 
observation of the excited $b$ baryon $\Xi_b^{*0}$. We also present
the observation of two decay modes of the $B_c$ meson and production properties of the $\Lambda_b$ baryon. 
All measurements were performed 
with the CMS experiment using pp collision data collected in 2011 at a center-of-mass energy of 7 TeV.

CMS is a general-purpose experiment at the Large Hadron Collider\,\cite{CMS}.
The inner detector contains a silicon tracker
composed of pixel layers at radii less than 15\,cm and strip layers out to a radius of
110\,cm.  The central region $|\eta|<1$ has 3 layers of pixels and 10 layers of
strips.  The cylindrical geometry of the central
region changes to disks in the $r-z$ plane for the forward region.  Each side of the
interaction region contains two endcap pixel layers and up to 12 layers of strips.
The tracker, PbWO$_4$ electromagnetic calorimeter, and
brass-scintillator hadron calorimeter are all immersed in a 3.8~T axial magnetic field.
Muons and hadrons are tracked within the pseudorapidity region $|\eta| < 2.4$, with
a $p_T$ resolution of about 1.5\% for tracks used in the analyses presented here.

\section{First observation of $\Xi_b^{*0}$}

\begin{figure}[h]
\begin{center}
\includegraphics[clip,width=0.45\linewidth]{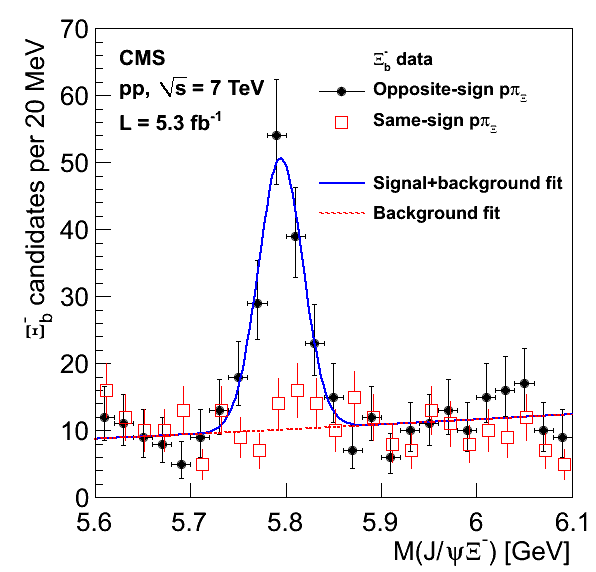}
\includegraphics[clip,width=0.45\linewidth]{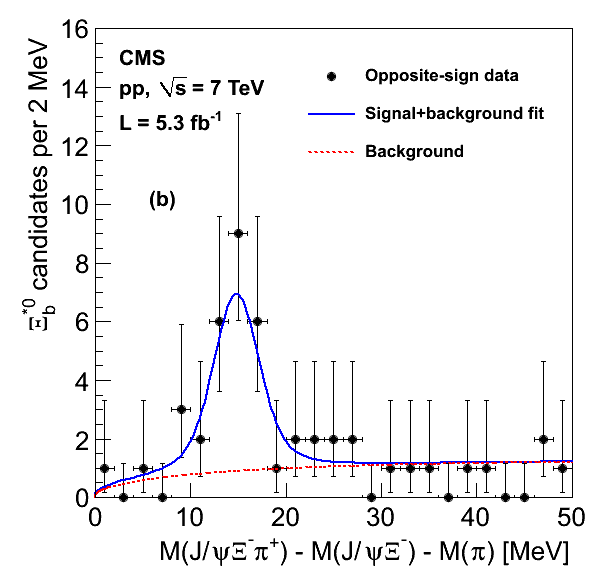}
\caption{Invariant-mass distributions for $\Xi_b^{-}$ candidates (left) and
$\Xi_b^{*0}$ candidates (right). The dashed (solid) curve represents the background 
(signal-plus-background) fit function.}
\label{fig:Xib}
\end{center}
\end{figure}

The first observation of a new beauty baryon,  $\Xi_b^{*0}$, is reported in decays
to $\Xi_b^{-}\pi^+$~\cite{Xib}. 
The precision of the CMS tracker is well utilized in this decay chain,
reconstructed with three secondary vertices. First, $\Lambda$ candidates are formed from
oppositely charged proton and pion tracks. The $\Lambda$ candidates are combined with
pion candidates, where the charge of the pion is same as that of the pion in the $\Lambda$
candidate, to form $\Xi^-$ candidates. The $\Xi^-$ candidates are combined with \JPsi\
candidates, which are formed from the combination of two oppositely
charged muons, to form $\Xi_b^{-}$ candidates. The muons are also used to trigger the event. 
Finally, the $\Xi_b^{-}$ candidates are combined with pion candidates to form $\Xi_b^{*0}$
candidates. A total of thirty variables are used in the final selection, with the most
signal-to-background discriminating power coming from displaced-track impact parameter
and displaced-vertex requirements. Figure~\ref{fig:Xib} shows the final invariant-mass 
distributions for the  $\Xi_b^{-}$ and  $\Xi_b^{*0}$ candidates. The significance of
the signal as measured by the log of the ratio of the fit likelihood modeled with and without
a signal is 6.9 standard deviations.

\section{$\Lambda_b$ production}

\begin{figure}[h]
\begin{center}
\includegraphics[clip,width=0.45\linewidth]{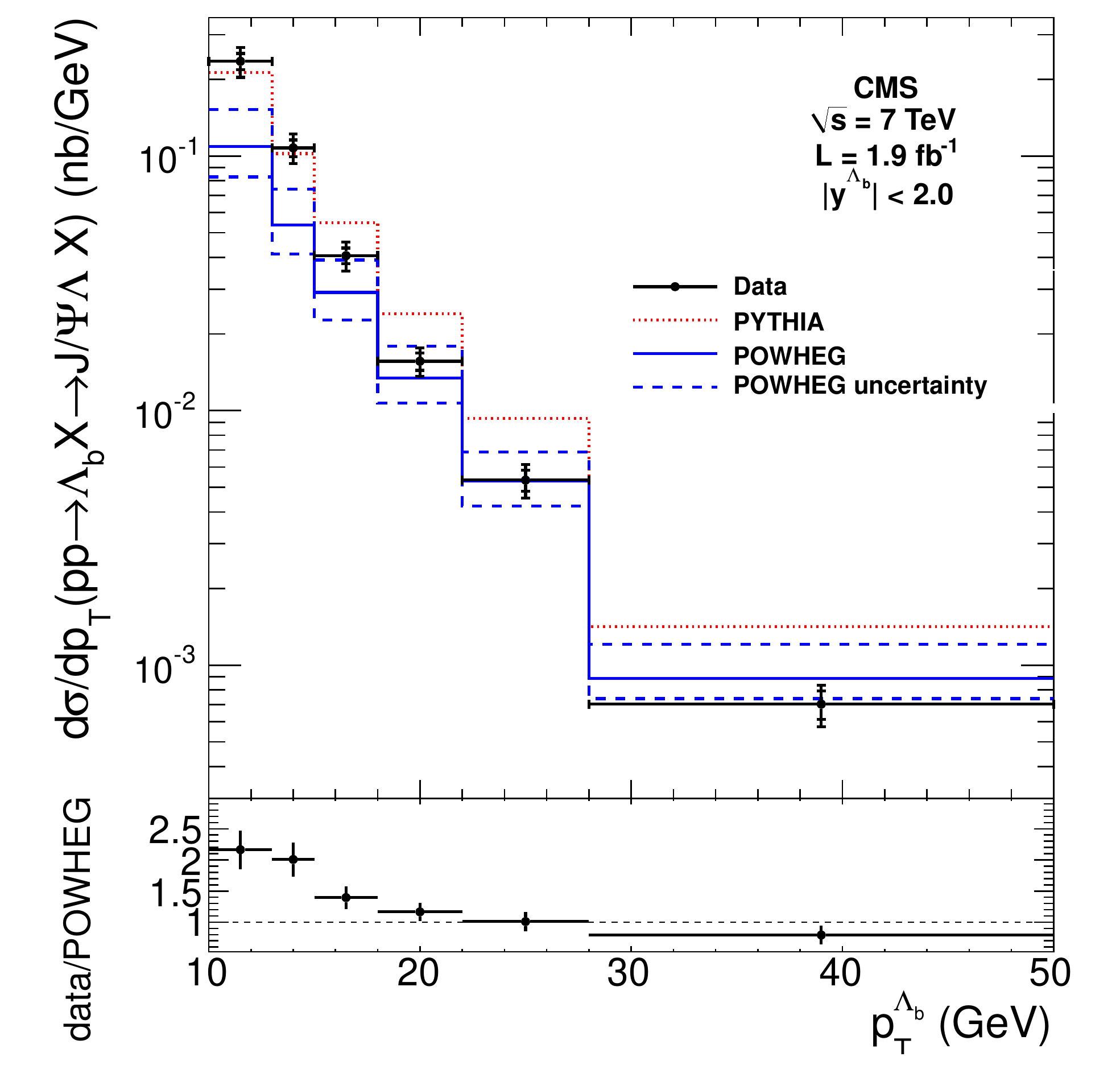}
\includegraphics[clip,width=0.45\linewidth]{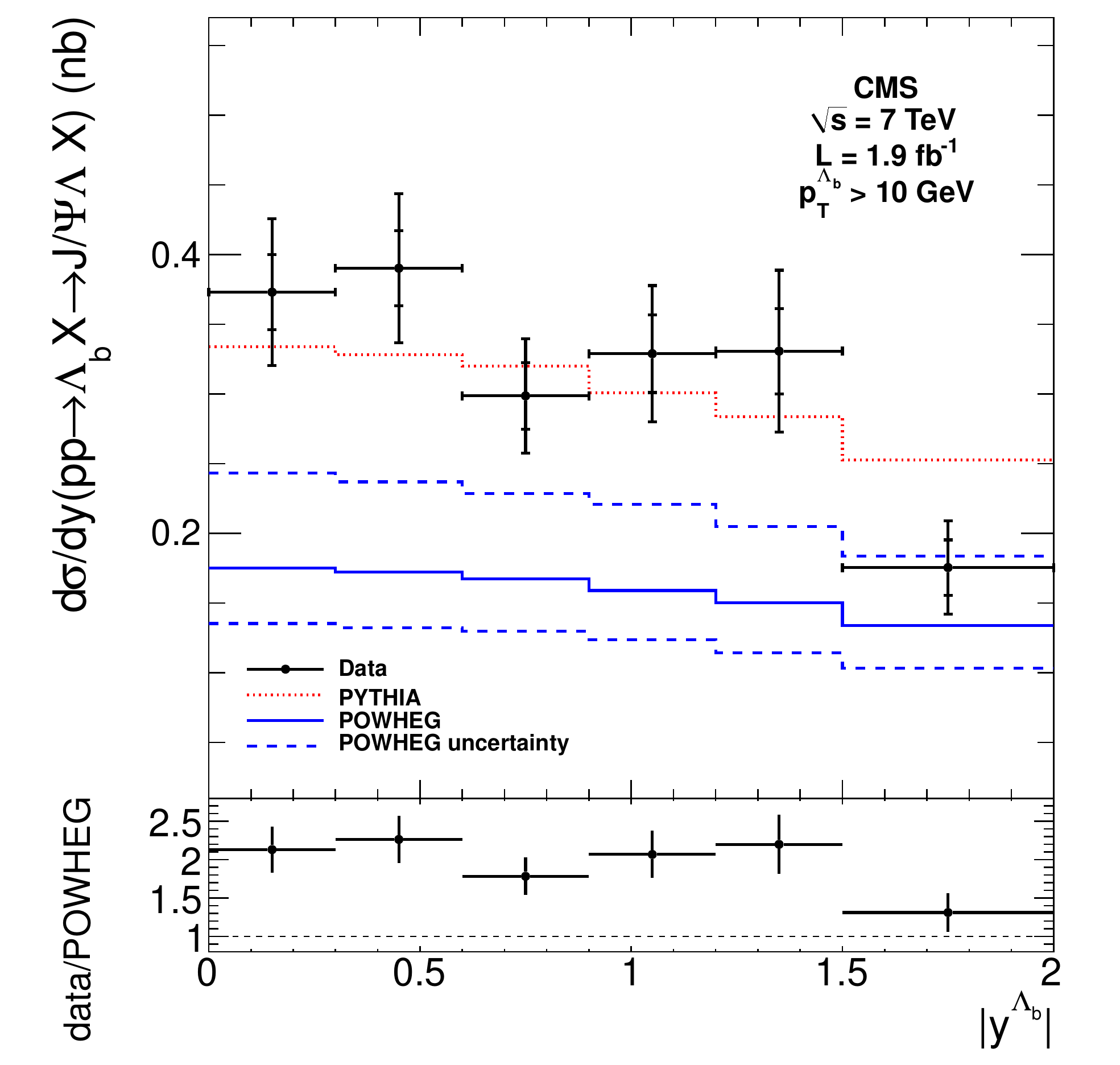}
\caption{
Measured differential cross sections times branching fraction $\dsdptbf$ (left)
and $\dsdybf$ (right) compared to the
theoretical predictions from PYTHIA and POWHEG, including theoretical uncertainties on the POWHEG predictions,
are shown in the top plots. Ratios of
data over the POWHEG predictions are shown in the bottom plots.}
\label{fig:Lb}
\end{center}
\end{figure}

Decays of \PgLb\ baryons to the final state $\JPsi\Lambda$, with \psimumu\ and \lamppi, are used
to measure the differential production cross section verses rapidity and transverse 
momentum~\cite{LbXsec}. 
Events are triggered with a displaced dimuon trigger that required dimuon $p_T > 7$ GeV and
a separation of the dimuon vertex from the beamspot greater than three times its uncertainty.
Oppositely charged muon candidates are combined to form \JPsi\ candidates, while $\Lambda$
candidates are formed from oppositely charged pion and proton candidates. The \PgLb\
candidates are formed by combining a \JPsi\ candidate with a $\Lambda$ candidate. The \PgLb\
yields are obtained by fitting the invariant-mass distribution to a peak above a combinatorial
background. The efficiency to reconstruct a \PgLb\ candidate is determined from a 
combination of techniques using Monte Carlo simulation and measurements in the CMS data, with a total
efficiency ranging from $0.3\%$ for low-$p_T$ \PgLb\ candidates to $4.0\%$ for high-$p_T$ 
candidates. The yields and efficiencies are computed separately in bins of \PgLb\
rapidity and $p_T$ to obtain a measurement of the differential production cross sections.
The production cross section is determined for $p_T$ of the \PgLb\ starting 
at 10 GeV and for the central region with the \PgLb\ rapidity extending up to 2.0. 
The results are shown in Fig.~\ref{fig:Lb}.

\section{Observation of $B_c$ Decay Modes}

\begin{figure}[h]
\begin{center}
\includegraphics[clip,width=0.45\linewidth]{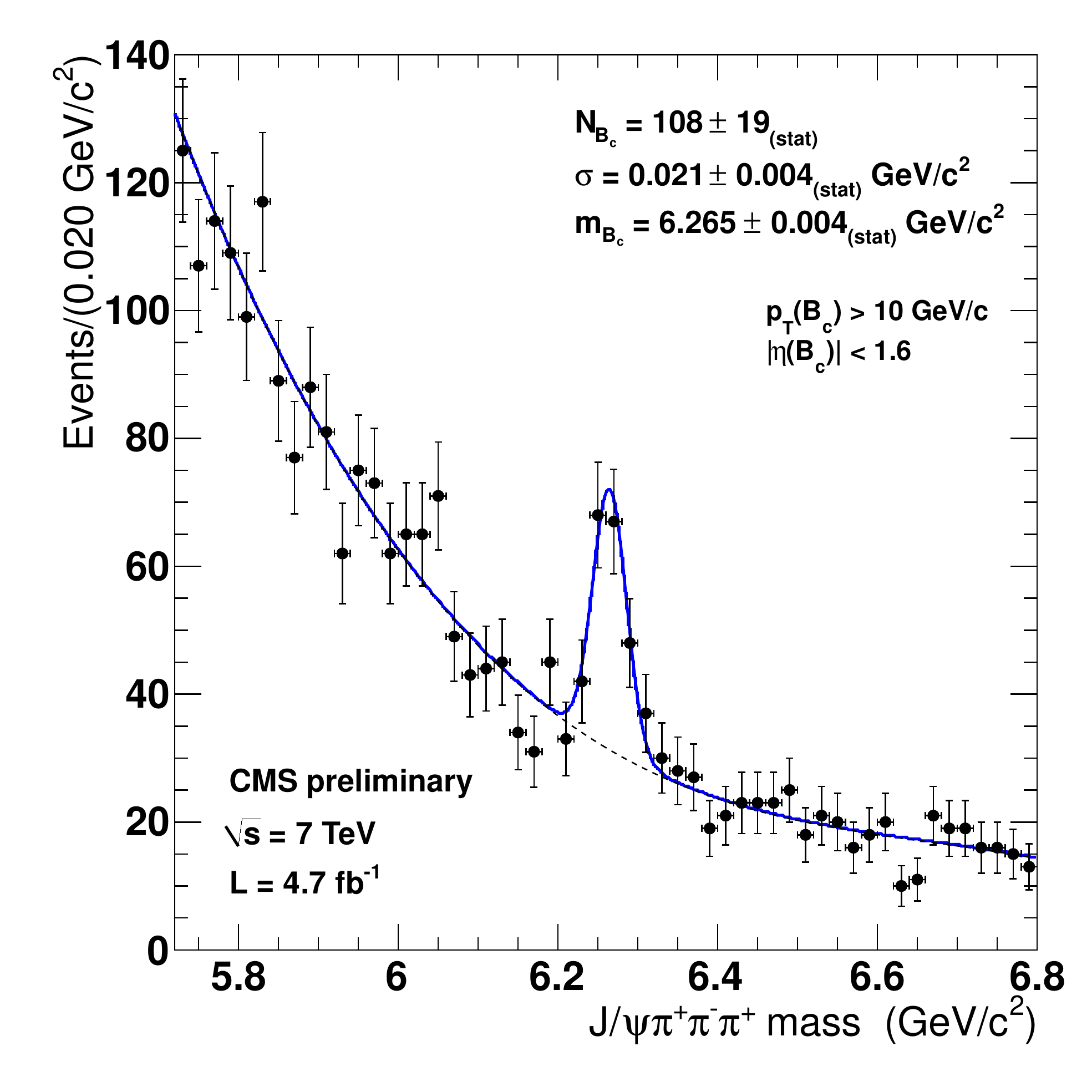}
\includegraphics[clip,width=0.45\linewidth]{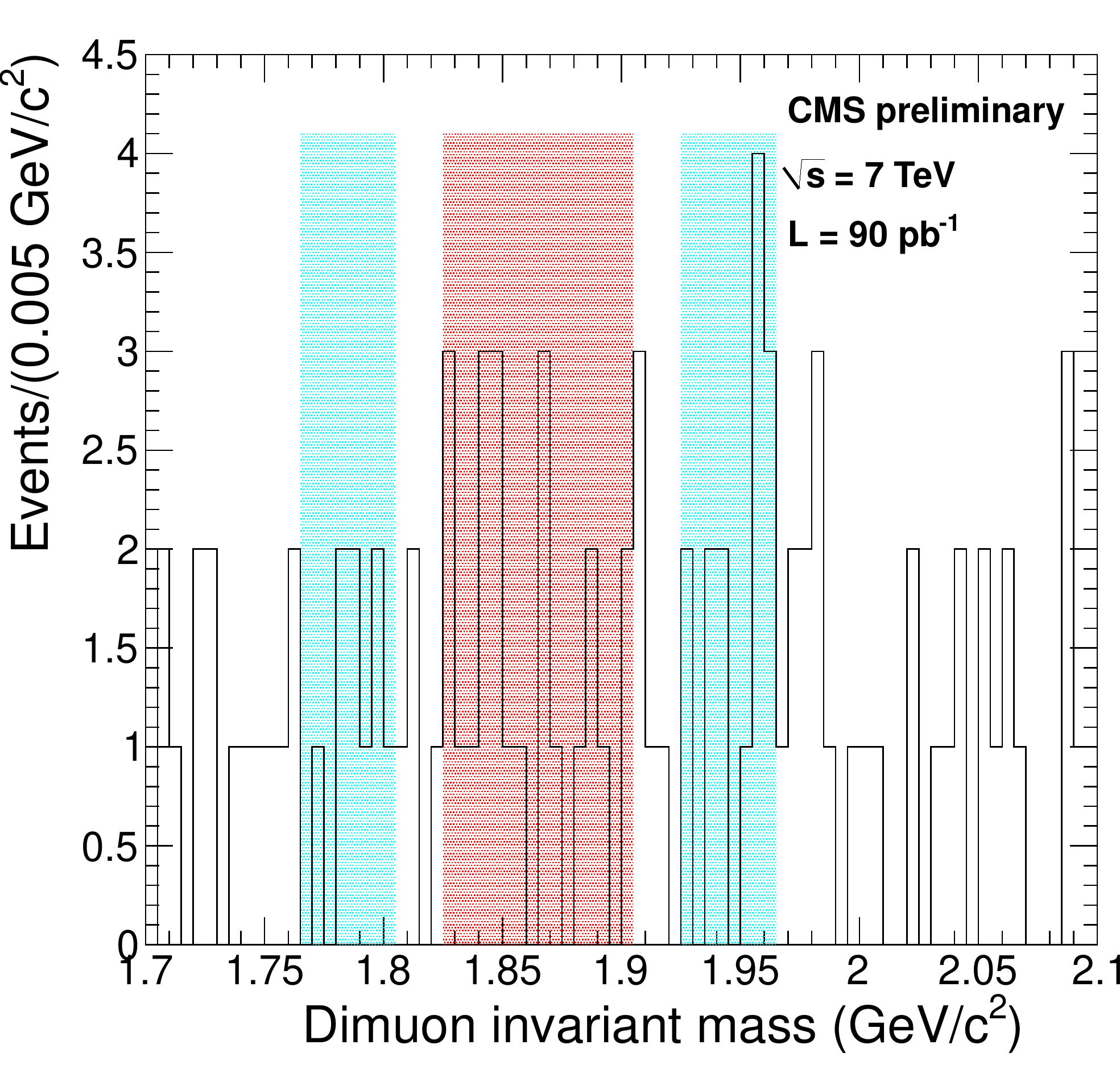}
\caption{Invariant-mass distribution for $B_c$ candidates in the $\JPsi\pi^+\pi^-\pi^+$
channel (left), where the dashed (solid) line represents the background (signal-plus-background) fit function,  
and $D^0\rightarrow\mu^+\mu^-$ candidates (right), where the signal (control) region is shown shaded in red (blue).}
\label{fig:BcD0}
\end{center}
\end{figure}

The $B_c^+$ meson, consisting of a beauty quark and a charm quark, presents a unique laboratory
to study heavy-quark dynamics. We present the observation of two decay channels,
$B_c^+\rightarrow\JPsi\pi^+$ and $B_c^+\rightarrow\JPsi\pi^+\pi^-\pi^+$~\cite{BcPAS}. 
In both channels,
\JPsi\ decays to two muons are used to trigger the events. Either one or three charged pion candidates
are combined with the \JPsi\ candidate to form $B_c^+$ candidates. A total yield of
330 $\pm$ 36 events are observed for the one-pion-decay channel, while $108\pm 19$ events are
observed for the three-pion-decay channel. Figure~\ref{fig:BcD0} (left) shows the invariant-mass 
distribution for the three-pion-decay channel.

\section{Search for $D^0\rightarrow\mu^+\mu^-$}

The rare flavor-changing neutral-current decay $D^0\rightarrow\mu^+\mu^-$ is highly suppressed
in the Standard Model, but can be enhanced in new-physics scenarios. We present results
from the first search for this decay at CMS based on a small portion of the total data set
representing $\approx$90 pb$^{-1}$~\cite{D0PAS}. 
A single-muon trigger is used to search for the signal
decay and for a control channel, $D^0\rightarrow K^-\mu^+\nu$, which has similar kinematics. 
Many systematic effects cancel in the ratio of the efficiencies for the signal and control channels.
Signal candidates are reconstructed by combining two oppositely
charged muon candidates. The dimuon candidate momentum is required to point back to the 
beamline. The control channel is reconstructed by identifying $D^0$ candidates from 
$D^{*+}\rightarrow D^0\pi^+$ decays. Backgrounds for the search sample are measured by 
interpolating from the dimuon mass sidebands in data on either side of the $D^0$ mass range.
No evidence of $D^0\rightarrow\mu^+\mu^-$ decays is
observed, and the control sample yield and known branching fraction are used to determine
an upper limit on the branching fraction. At $90\%$ confidence level, an upper limit on the $D^0\rightarrow\mu^+\mu^-$
branching fraction is calculated with the $\rm{CL_s}$ technique as $< 5.4 \times 10^{-7}$.
Figure~\ref{fig:BcD0} (right) shows the invariant-mass distribution for the $D^0\rightarrow\mu^+\mu^-$
candidates, where the red and blue regions are the signal and control regions, respectively.

\section{Search for $B^0_{(s)}\rightarrow\mu^+\mu^-$}

The rare flavor-changing neutral-current decays $B^0_{(s)}\rightarrow\mu^+\mu^-$ are highly suppressed
in the Standard Model, but can be altered in many new-physics scenarios. In particular, limits at the
LHC are now approaching the Standard
Model prediction for $B^0_s\rightarrow\mu^+\mu^-$ of $(3.2\pm0.2)\times10^{-9}$~\cite{BmmPred}. 
We present the 
results of a simultaneous search for $B^0\rightarrow\mu^+\mu^-$ and $B^0_s\rightarrow\mu^+\mu^-$ at
CMS in $pp$ collisions at $\sqrt{s}=7\,\mathrm{TeV}$, with a data sample corresponding to an
integrated luminosity of $5\,\mathrm{fb}^{-1}$~\cite{BmmPaper}. Candidate $B^0_{(s)}$ mesons are reconstructed
by fitting oppositely charged muon candidates to a common vertex. Signal events are selected
preferentially by requiring the dimuon vertex to be well separated from the beamspot, the dimuon
candidate to be well isolated from other detector activity and that the two muons form a good vertex.

\begin{figure}[h]
\begin{center}
\includegraphics[clip,width=0.40\linewidth]{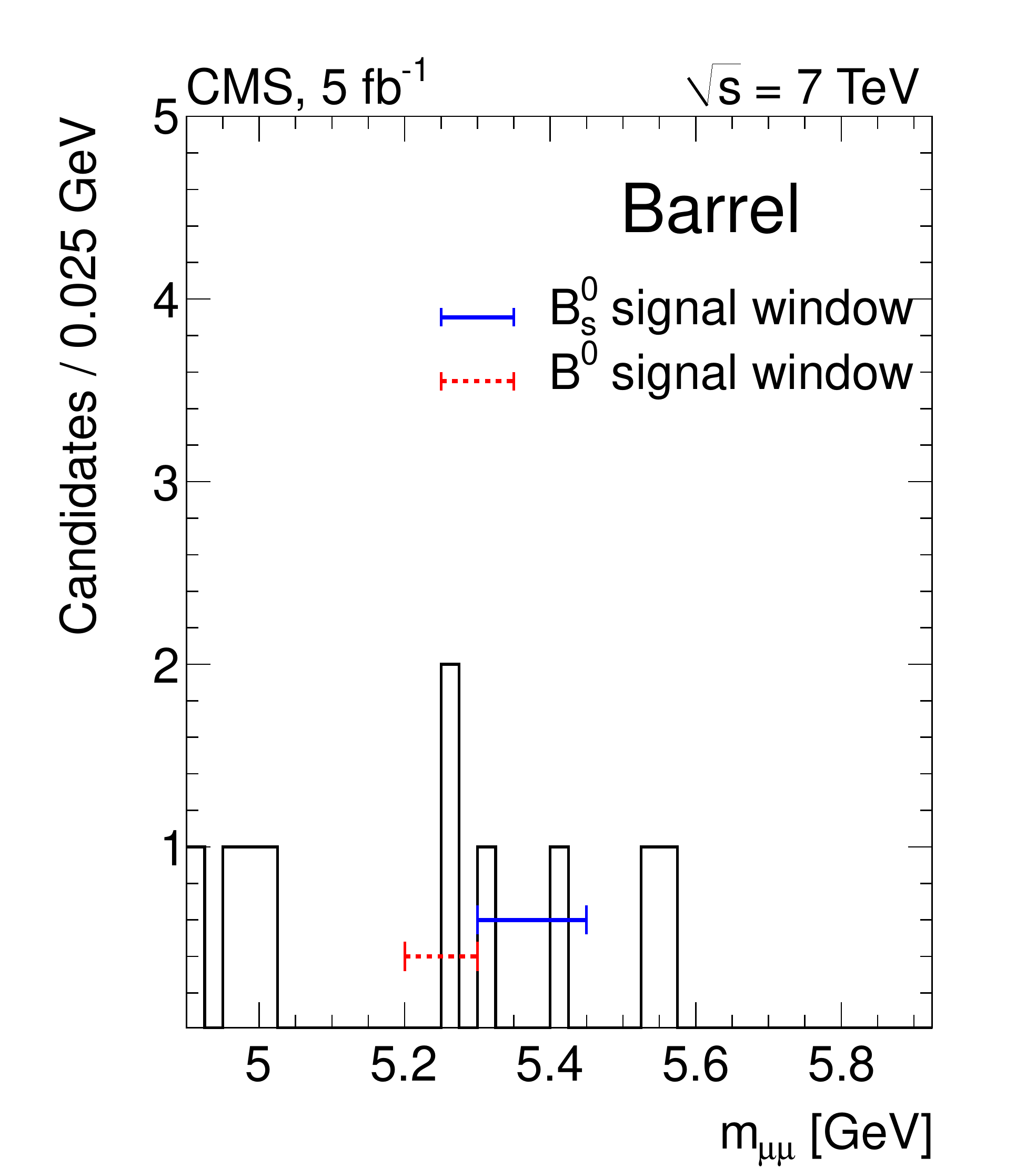}
\includegraphics[clip,width=0.40\linewidth]{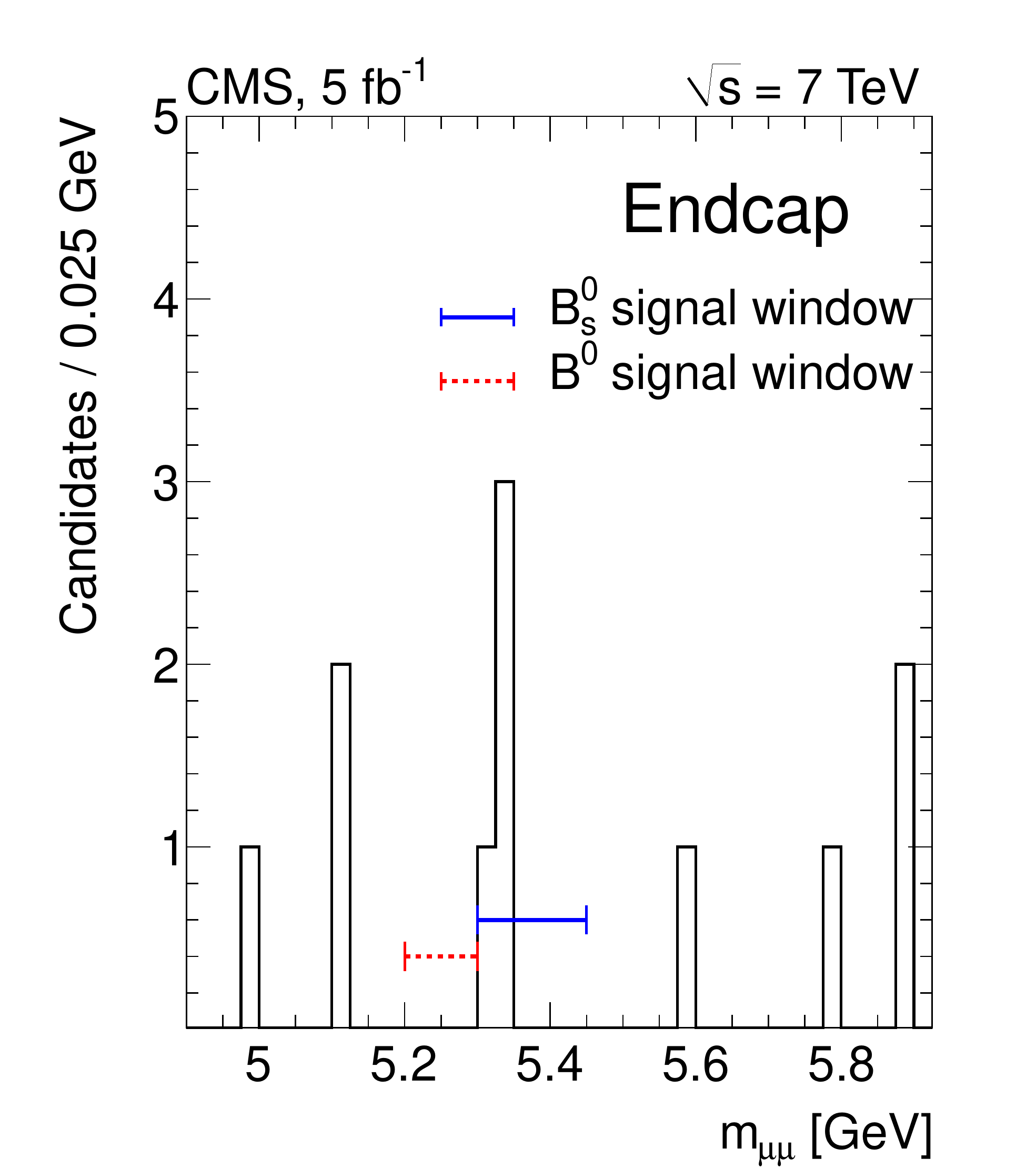}
\includegraphics[clip,width=0.60\linewidth]{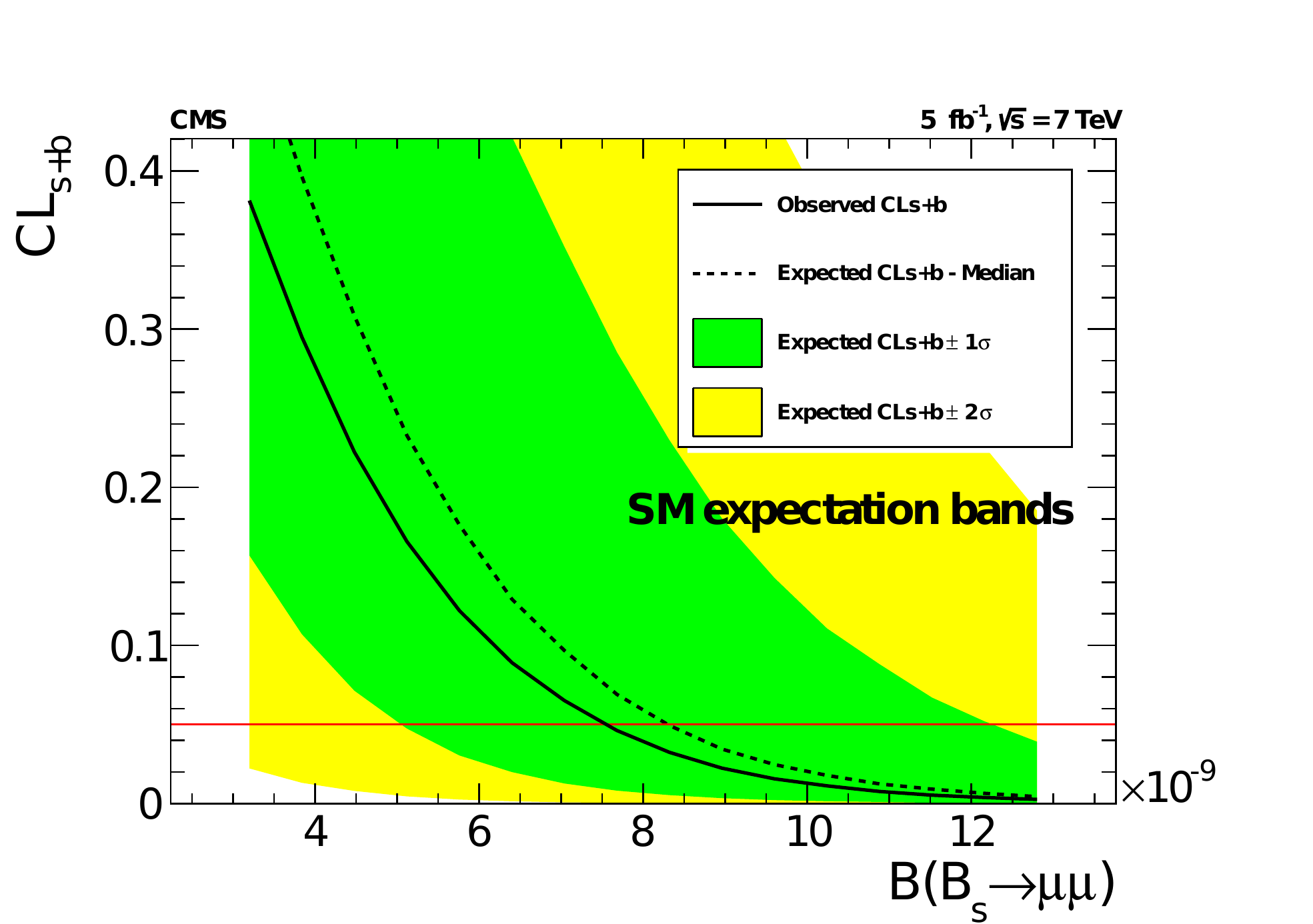}
\caption{
Invariant-mass distributions for $B^0_{(s)}\rightarrow\mu^+\mu^-$ candidates in
the barrel (upper left) and endcap (upper right) regions. The blue and red regions are the signal windows for
$B^0_s$ and $B^0$, respectively. The observed (solid line) and expected (dotted line) $\rm{CL_{s+b}}$ curves
for $B^0_{s}\rightarrow\mu^+\mu^-$ compared to SM expectation (bottom). The green and yellow regions show
the $\pm 1$ and $\pm 2$ standard deviation variations in the expected sensitivities.}

\label{fig:Bmm}
\end{center}
\end{figure}

The search is performed by counting events with dimuon invariant mass within appropriate windows
for $B^0_{(s)}$ mesons. The event sample is divided into two categories, barrel and endcap, where
the former contains only candidates when both muons are within $|\eta| < 1.4$, which have the best momentum
resolution. Backgrounds arise from combinatorial muon pairs, which are measured by extrapolating
from the sidebands of the dimuon invariant-mass distributions, and from peaking backgrounds from
semileptonic or hadronic $B$ decays where hadrons are misidentified as muons. The peaking background is
measured by scaling the background expectation from simulation by the hadron misidentification rates
measured in CMS data. No significant signal is observed in either channel. The dimuon invariant-mass 
distributions are shown in the top plots of Fig.~\ref{fig:Bmm}.

Upper limits on the branching fractions are determined through a normalization channel, 
$B^+\rightarrow\JPsi K^+$. Many systematic uncertainties cancel in the ratio of the yields
and efficiencies between the signal and normalization channels. Upper limits of
${\calB}(\mathrm{B}_\mathrm{s}^0\to\mu^+\mu^-) < 7.7\times10^{-9}$ and ${\cal
B}(\mathrm{B^0}\to\mu^+\mu^-) < 1.8\times10^{-9}$ are set at $95\%$ confidence level using the
$\rm{CL_s}$ technique. The $\rm{CL_{s+b}}$ curve for $B^0_s\rightarrow\mu^+\mu^-$ is shown in
the lower plot of Fig.~\ref{fig:Bmm}. Additionally,
a combination of the CMS result with measurements of the same channels from LHCb and ATLAS
is performed, which provides the best upper limits to date for these channels, 
${\calB}(\mathrm{B}_\mathrm{s}^0\to\mu^+\mu^-) < 4.2\times10^{-9}$ and 
${\calB}(\mathrm{B^0}\to\mu^+\mu^-) < 8.1\times10^{-10}$ at $95\%$ confidence level~\cite{BmmCombination}.

\section{Conclusion}

A selection of results from the CMS heavy-flavor physics program has been shown, including new limits on searches
for rare decays and new measurements of heavy-flavor production.

\end{document}